\documentclass[twocolumn,showpacs,amsmath,amssymb,pra]{revtex4}

\usepackage{graphicx}
\usepackage{dcolumn}
\usepackage{bm}
\usepackage{bbm}
\usepackage{amsfonts}

\newcommand{\ve}[1]{\mathbf{#1}}
\newcommand{\bra}[1]{\langle #1|}
\newcommand{\ket}[1]{|#1\rangle}

\begin{document}

\title{Scaling of running time of quantum adiabatic algorithm for propositional satisfiability}
\author{Marko \v Znidari\v c}
\affiliation{
Physics Department, Faculty of Mathematics and Physics,University of Ljubljana, SI-1111 Ljubljana, Slovenia}
\affiliation{Department of Quantum Physics, University of Ulm, D-89069 Ulm, Germany}

\begin{abstract}
We numerically study quantum adiabatic algorithm for the propositional satisfiability. A new class of previously unknown hard instances is identified among random problems. We numerically find that the running time for such instances grows exponentially with their size. Worst case complexity of quantum adiabatic algorithm therefore seems to be exponential.
\end{abstract}


\date{\today}

\maketitle

\section{Introduction}
Computers play a vital role in modern society. Since its inception in the middle of the previous century the power of digital computers has been growing exponentially with time. How long can this growth be sustained? Great interest in recent years in quantum computing is partially fuelled by the discovery that quantum computers could perform certain tasks faster than any classical computer. An example is the famous Shor factoring algorithm~\cite{Shor:97} which is polynomial, whereas the best known classical algorithm is super-polynomial. According to computational complexity, problems can be divided into two large groups. Those for which the time to find a solution grows polynomially with the size of the problem belong to the so-called P class (polynomial) and those that require polynomial time to verify the solution belong to NP (non-deterministic polynomial). Especially important subset of NP problems is called NPC (NP-complete). They have the property that any NP problem can be transformed to NPC problem in a polynomial time. Therefore, finding a polynomial algorithm for a single NPC problem would immediately provide a polynomial algorithm for all NP problems. Currently all known algorithms need exponential time to solve NPC problems. In a vague way it can be said that NPC are the hardest of NP problems.
\par
Recently a novel way of doing quantum computation via adiabatic evolution has been suggested~\cite{Farhi:00,Farhi:01}. The idea of using adiabatic evolution to do quantum computation is very simple and elegant. One starts with the system in the ground state of the initial Hamiltonian $H(0)$. Then the Hamiltonian is adiabatically slowly changed from $H(0)$ to the final $H(1)$, whose ground state encodes the solution to the problem we want to solve. The adiabatic theorem then ensures that if the changing of the Hamiltonian is sufficiently slow, we end up in the ground state of $H(1)$ at the end, thereby obtaining the solution to our problem. Numerical simulation of the adiabatic algorithm for a NPC problem called exact cover~\cite{Farhi:01} indicated that quantum adiabatic algorithm might need running time that grows only quadratically with the size of the problem. Subsequently, there have been many studies of quantum adiabatic algorithms, mostly numerical~\cite{Smelyansky:01,Childs:02,Hogg:03,Mitchell:04} but also some rigorous results are known~\cite{vanDam:01,Roland:02,Reichardt:04,Roland:04}. It has also been shown recently that adiabatic computation is polynomially equivalent to standard quantum computation~\cite{aharonov:04}. Adiabatic algorithm has also been experimentally realized on a NMR quantum computer~\cite{Steffen:03}. While numerics for number partitioning problem showed exponential scaling~\cite{Smelyansky:01} the scaling of running time in other NPC problems is still unclear. For instance, for a paradigmatic example of a NPC problem called 3-SAT  (3-satisfiability), polynomial scaling $\sim n^3$ of median cost has been found~\cite{Hogg:03}. Needless to say, the implications of having a polynomial quantum adiabatic algorithm for NPC problem would be enormous. But we have to keep in mind that the computational complexity is defined in terms of the worst case performance.
\par
 Although the studies so far focused on presumably hard instances of 3-SAT problems we will show that there exists a class of even harder 3-SAT instances not known before. We will present a clear numerical evidence for an exponential scaling of running time of quantum adiabatic algorithm for these 3-SAT instances. This finding could also be relevant for classical algorithms development, where hard instances are used in algorithm design and testing.
\par
The outline of the paper is as follows. In Section~\ref{sec:3sat} we introduce a problem studied, namely a random 3-SAT. In Section~\ref{sec:qaa} quantum adiabatic algorithm is defined and the degeneracies of the initial and final Hamiltonian are explored. In Section~\ref{sec:lz} we study the probability to successfully obtain correct result. Then in Section~\ref{sec:spec} the energy spectrum during adiabatic evolution is studied and finally in Section~\ref{sec:gap} the scaling of the energy gap and running time is presented.

\section{Random 3-SAT}
\label{sec:3sat}
In the present paper we will consider 3-SAT problem. It is a paradigmatic example of a NPC problem. 3-SAT formula is a logical statement involving $n$ boolean variables $b_i$. It consists of $m$ clauses $C_i$ in conjunction (logical AND $=\land$)
\begin{equation}
C_1 \land C_2 \land \cdots \land C_m,
\label{eq:3sat}
\end{equation}
and each clause $C_i$ is a disjunction (logical OR $=\lor$) of 3 literals, where a literal is a variable $b_i$ or its negation $\lnot b_i$ (logical NOT $=\lnot$). To illustrate, an instance of a 3-SAT formula with $n=4$ variables and $m=2$ clauses is 
\begin{equation}
(b_2 \lor \lnot b_3 \lor b_4) \land (b_1 \lor b_2 \lor \lnot b_3).
\label{eq:example}
\end{equation}
The problem is to decide whether a given 3-SAT formula is satisfiable, i.e. whether there exists a prescription of variables $b_i$ such that 3-SAT formula is true. Such prescription is called a solution, the number of which will be denoted by $r$. The formula given as an example~(\ref{eq:example}) has many solutions, one being for instance $(b_4b_3b_2b_1)=1101$, where $1$ and $0$ denote true and false, respectively. 
\par
Formula that is a conjunction of disjunctions is said to be in a conjunctive normal form (CNF). Any logical statement consisting of $\land$, $\lor$ and $\lnot$ operators can be rewritten in a CNF form. While 2-SAT problem (CNF formula having at most $2$ literals in each clause) belongs to P, 3-SAT is in NPC. Besides being paradigmatic example of a NPC problem it has wide range of applicability in e.g. scheduling problems, hardware verification etc... It is also directly related to deductive reasoning important in artificial intelligence. If we are given a set $\Sigma$ of facts (statements) $C_i$, $\Sigma=\cup C_i$, a new statement $C_{\rm new}$ can be deduced iff a union $\Sigma \cup \{\lnot C_{\rm new} \}$ is not satisfiable (we arrive at contradiction assuming $\lnot C_{\rm new}$), i.e. we have to solve a SAT problem. 
\par
As the computational complexity is defined in terms of worst case performance, hard instances of 3-SAT have been especially intensely studied. Frequently they are obtained by randomly drawing clauses, so-called random 3-SAT. A random 3-SAT consists of $m$ different random clauses, where each clause is obtained by picking $3$ different variables $b_i$ and negating each with probability $1/2$. This is the procedure we used to obtain random instances of 3-SAT. As we were mainly interested in formulae with only one solution, we solved each instance and rejected those not having exactly one solution. In the literature on the other hand they usually study random 3-SAT formulae with an arbitrary number of solutions. In such case a phase-transition is found~\cite{Kirkpatrick:94} with the hardest instances occurring around the phase-transition point of $m/n \approx 4.2$. As we will see, random 3-SAT instances with exactly one solution and small $m/n$ will turn out to be harder than those at the phase-transition (and having arbitrary number of solutions). Correct choice of hard instances of 3-SAT is therefore absolutely essential in order to see clear exponential scaling of running time. For more information about a phase-transition in random 3-SAT see also papers in the~\cite{volumeAI} and also~\cite{monasson:99,cocco:02}. For classical SAT problem solving see e.g. collection~\cite{SAT}.

\section{Hamiltonian for adiabatic algorithm}
\label{sec:qaa}
\begin{figure}[!h]
\centerline{\includegraphics[width=3.3in]{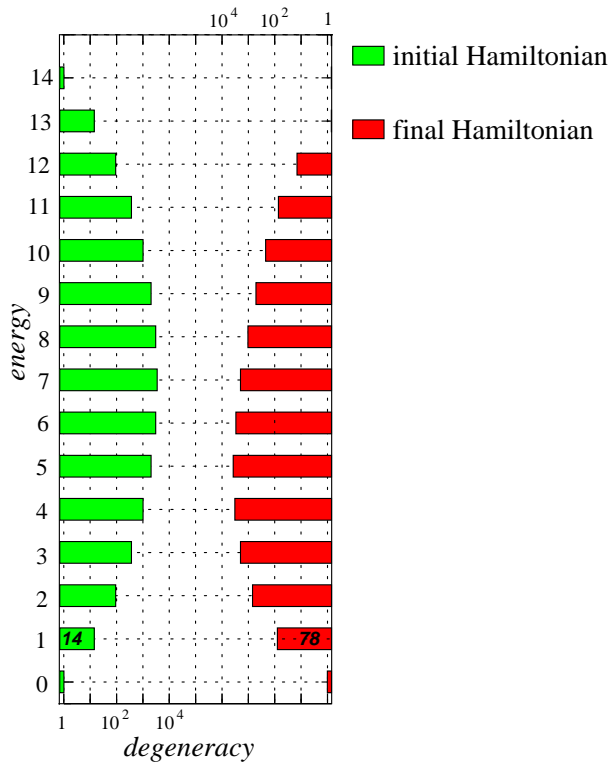}}
\caption{(Color online)~Degeneracy of the initial (left) and final Hamiltonian (right) for one 3-SAT instance with $n=14$, $m/n=3$ and exactly one solution $r=1$. Lower part of spectrum for all intermediate times $s$ is shown in Fig.~\ref{fig:test14}.}
\label{fig:deg14}
\end{figure}

 The construction of Hamiltonian for the 3-SAT problem is straightforward. For each variable we have to have available two states, i.e. one qubit, giving the dimension of the total Hilbert space $N=2^n$. The initial $H(0)$ is problem independent and we choose it to be a sum of one-qubit Hamiltonians $H_i$ acting on $i$-th qubit,
\begin{equation}
H(0)=\frac{1}{2}\sum_{i=1}^{n}{H_i\otimes \mathbbm{1}},\qquad H_i=\begin{pmatrix}
1 & -1 \cr
-1 & 1
\end{pmatrix}.
\label{eq:H0}
\end{equation}
The ground state $\ket{E_0(0)}$ of the initial Hamiltonian has energy $E_0=0$ and is a uniform superposition of all computational states,
\begin{equation}
\ket{E_0(0)}=\frac{1}{\sqrt{N}}\sum_{b_1,\ldots,b_n=0,1}{\ket{b_n\ldots b_1}},
\label{eq:psi0}
\end{equation}
where a label of state $\ket{b_{n}b_{n-1}\ldots b_1}=\ket{\ve{b}}$ denotes a binary expansion of the state, i.e. the value of each qubit. In fact the whole energy spectrum of the initial Hamiltonian is easily calculated. It consists of integer energies, $E_i(0)=i$, with the degeneracy of level $E_i(0)$ being equal to ${n \choose i}$. The final Hamiltonian $H(1)$ is problem dependent. We use a diagonal Hamiltonian in the computational basis, with the energy of state $\ket{\ve{b}}$ equal to the number of clauses it violates. As all states violating a given clause have the same values of $3$ variables occurring in that clause, each clause can be represented by a single 3-qubit gate. For our example (\ref{eq:example}) we would have two terms,
\begin{equation}
H(1)=\ket{010}\bra{010}_{432}\otimes \mathbbm{1} +\ket{100}\bra{100}_{321}\otimes \mathbbm{1},
\label{eq:H1example}
\end{equation}
where subscripts denote on which qubits the operator acts. Therefore, a state $\ket{\ve{b}}$ satisfying all clauses (i.e. a solution) would have energy $0$, state violating a single clause has energy $1$ and so on. Energy spectrum of $H(1)$ is therefore composed of integer values between $E_0(1)=0$ for the ground state and $E_m(1)=m$ for a state that would violate all clauses. Of course, the degeneracy $r$ of the state $E_0(1)$ of $H(1)$, i.e. the number of solutions, depends on the particular instance in question. In the thermodynamic limit, $n \to \infty$, 3-SAT formulae with $m/n$ below a ``phase transition'' point at $m/n \approx 4.2$ have many solutions, whereas formulae with $m/n>4.2$ have no solution~\cite{Kirkpatrick:94}. Whereas the spectrum of $H(0)$ is fixed, the spectrum of $H(1)$ is instance dependent. An example of spectrum of $H(0)$ and $H(1)$ for one particular 3-SAT instance having $n=14$ variables and $m=42$ clauses is shown if Fig.~\ref{fig:deg14}. From the spectrum of $H(1)$ we see for instance, that there are no states violating more than $12$ clauses and the most abundant are $3720$ states violating $5$ clauses (with energy $E=5$). There is only one state with energy $E=0$, thus there is only one solution. Throughout the work we will focus only on 3-SAT problems having exactly one solution, $r=1$. Later we will argue that such problems are expected to be the hardest.
\par
The interpolating Hamiltonian between $H(0)$ and $H(1)$ is chosen according to the following prescription,
\begin{equation}
H(s)=(1-s) H(0)+ s H(1),\qquad s=\frac{t}{T},
\label{eq:Hs}
\end{equation} 
where $s=t/T$ is a dimensionless time and $T$ is the total running time of the adiabatic algorithm. The evolution of an arbitrary state $\ket{\psi}$ is given by a time dependent Schr\" odinger equation
\begin{equation}
\frac{1}{(dt/ds)}{\rm i} \hbar \frac{d}{ds} \ket{\psi}=H(s) \ket{\psi},
\label{eq:schrodinger}
\end{equation}
where for our choice of constant speed along the interpolating path $H(s)$ (\ref{eq:Hs}), $(dt/ds)$ is constant and equal to $T$. We set $\hbar=1$ throughout the paper.
\par
The interpolating ``path'' between $H(0)$ and $H(1)$ is quite arbitrary as well as the initial Hamiltonian $H(0)$, while the final Hamiltonian is determined by the problem in question. Instead of having uniform speed of interpolation we could vary it according to the energy gap. Such refinements are not the subject of the present paper. Trough adiabatic quantum computation hard question of computational complexity is translated into (perhaps?) easier question of the scaling of the energy gap. At least there are plenty of tools available for studying energy gaps.

\section{Failure probability}
\label{sec:lz}
In studying quantum adiabatic algorithms one usually numerically looks at the probability of successfully finding the solution for different running times $T$~\cite{Farhi:01,Smelyansky:01,Childs:02,Hogg:03}. The adiabaticity condition, guaranteeing adiabatic evolution, is usually stated as 
\begin{equation}
T \gg \hbar \int_0^1{\frac{||dH/ds||}{\gamma(s)^2}ds},
\label{eq:adiabatic}
\end{equation}
where $\gamma(s)=E_1(s)-E_0(s)$ is the energy gap between the ground state and the first excited state. Adiabatic condition can also be stated locally, saying that the local inverse speed has to be larger than
\begin{equation}
\frac{dt}{ds} \gg \hbar \frac{||dH/ds||}{\gamma(s)^2}. 
\end{equation}
\par
The above two adiabatic conditions~(\ref{eq:adiabatic}) give us the necessary condition for the adiabatic evolution. What would be desirable to know is also the probability of non-adiabatic transitions. This would then give us a direct way to calculate necessary running time for the desired probability to stay in the ground state at the end. For $2$ and $3$ level systems there is an exact expression. In the limit of slow evolution it is the famous Landau-Zener formula~\cite{Zener:32,Landau:58}, giving the probability of a transition in a $2$ level system, where the two eigenenergies are $E_{0,1}(s)$,
\begin{eqnarray}
P_{\uparrow}&=&\exp{\left( -\frac{T}{\tau_{\rm LZ}} \right)},\qquad \tau_{\rm LZ}=\frac{2A\hbar}{\pi \Delta^2},\nonumber \\
E_{0,1}(s)&=&\pm \frac{1}{2} \sqrt{\Delta^2+(A s)^2},
\label{eq:LZ}
\end{eqnarray}
here $\Delta$ is a minimum gap $\gamma(s)$, $T$ is a constant parameter connecting $s=t/T$ and $t$ runs from $-\infty$ to $\infty$. For a discussion of transitions in multilevel systems see~\cite{Wilkinson:00}. Landau-Zener formula has been used before~\cite{Mitchell:04} to describe adiabatic algorithm under the assumption that random matrix statistics applies to avoided crossings. Two level transition probability is also used in the adiabatic algorithm for Grover search~\cite{Roland:02}.
\par
\begin{figure}[!h]
\centerline{\includegraphics[width=3.3in]{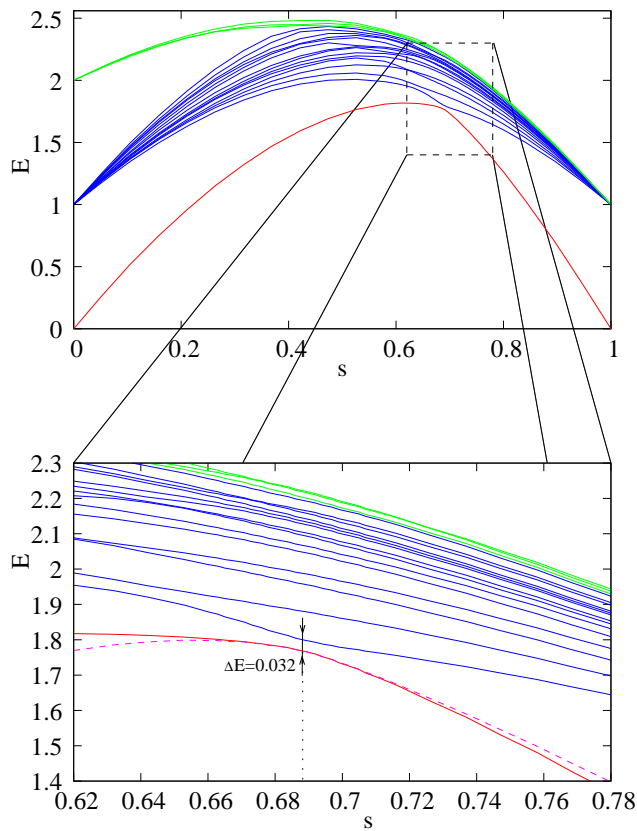}}
\caption{(Color online)~Lower part of spectrum (lowest 18 levels out of total $N=16384$) for one instance of 3-SAT with $n=14$, $m/n=3$, the same instance as in Fig.~\ref{fig:deg14}. We can see avoided crossing at $s \approx 0.7$. Dashed line is fitted avoided crossing, i.e. line $E_1(s)-\sqrt{\Delta^2+(A s)^2}$, whose parameters are obtained by fitting parabola at the crossing.}
\label{fig:test14}
\end{figure}
Let us first have a look at the lower part of a typical spectrum of $H(s)$ to see what is the nature of transitions. To find the lowest eigenvalues we used implicitly restarted Lanczos method~\cite{arpack}, suitable for sparse eigenvalue problems. By this we could find lowest few eigenvalues for $n$ upto $20$, i.e. Hilbert space sizes $N\sim 10^6$. One particular example for $n=14$ and $m=42$ is shown in Fig.~\ref{fig:test14}. Only lowest $18$ levels out of total $N=16384$ are shown. We can see that there is {\em only one avoided crossing}, the same was the case in all other cases we have checked. The reason to have only one avoided crossing is unclear to us. It might be connected with the fact that we have a finite gap at the beginning and at the end of the algorithm. Similar behaviour has been found by other researchers~\cite{Hogg:03}. Now provided we have only one close encounter of two lowest levels, we can use degenerate perturbation theory, i.e. consider only two closest levels. Transition probability is then given simply by the Landau-Zener formula (\ref{eq:LZ}). We first checked how well the Landau-Zener formula describes the transition probability for a real 3-SAT case, such as is for example the one in Fig.~\ref{fig:test14}.  
\begin{figure}[!h]
\centerline{\includegraphics[height=3.3in,angle=-90]{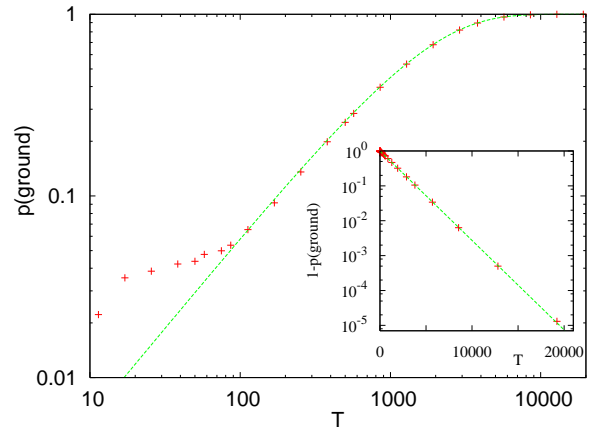}}
\caption{(Color online)~Comparison of success probability $p({\rm ground})$ obtained with direct numerical simulation (pluses) with $P_{\uparrow}$ obtained from Landau-Zener formula (\ref{eq:LZ}). In the inset the same data is shown in a semi-log scale for longer times. All is for the same 3-SAT instance shown in Fig.~\ref{fig:test14}.}
\label{fig:LZ}
\end{figure}
To obtain probability for a transition from the ground state using Landau-Zener formula (\ref{eq:LZ}) we fitted parabola around the avoided crossing to determine two parameters, the gap $\Delta$ and the asymptotic slope $A$, needed in the Landau-Zener formula. For a case in Fig.~\ref{fig:test14} for instance, we obtain $\Delta=0.0318$ and $A=2.67$ (dashed line in Fig.~\ref{fig:test14}). Then we compared $P_{\uparrow}$ (\ref{eq:LZ}) with the result of a direct numerical simulation of a time dependent Schr\" odinger equation (\ref{eq:schrodinger}). We discretized time into small steps $dt$ (typically $dt \sim 0.1$) and then calculated one-step propagator $U(dt)=\exp{(-{\rm i} H(t) dt/\hbar)}$ by expansion in a power series. The precision has been controlled throughout the calculation. At the end of the simulation, at time $t=T$, we obtain a final state $\ket{\psi(T)}$ and then calculate the overlap with the ground state, giving us the numerical probability $p({\rm ground})$ to remain in the ground state. In Fig.~\ref{fig:LZ} we compare this numerical value with the Landau-Zener formula. One can see a very good agreement already in a regime of small $T$, where the probability to stay in the ground state is small. Therefore, the Landau-Zener formula perfectly describes the probability to stay in the ground state in all practically relevant regime (i.e. for high $p({\rm ground})$). From now on we will focus on the scaling of $\tau_{\rm LZ}$ with the size $n$ of a problem.

\section{Spectrum}
\label{sec:spec}
\begin{figure}[!h]
\centerline{\includegraphics[height=3.3in,angle=-90]{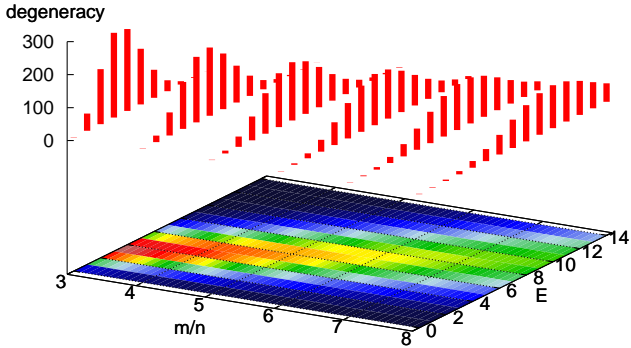}}
\caption{(Color online)~Degeneracies of the spectrum of final hamiltonian $H(1)$ for different ratios $m/n$. Degeneracy of the first excited state increases with decreasing $m/n$. All points are an average over $10000$ random 3-SAT instances with exactly one solution, i.e. the degeneracy of the ground state with $E=0$ is $r=1$.}
\label{fig:spekter_n10}
\end{figure}

There are two parameters determining $\tau_{\rm LZ}$. The asymptotic slope $A$ at the avoided crossing has value of around $2$ (ground state energy changes from $0$ to $\approx 1$ at the avoided crossing and back to $0$ at $s=1$) and does not vary appreciably with $n$. The main dependence of $\tau_{\rm LZ}$ on $n$ will be therefore given by the scaling of the minimal gap $\Delta$. Heuristically one could argue that the gap $\Delta$ will be smaller when more excited levels are crowded into a region of energies between $E\approx 1$ and $E\approx 2$, where the avoided crossing takes place. The number of such levels is connected with the degeneracies of the first excited levels at the beginning and at the end of the algorithm. The degeneracy of the first excited level of $H(0)$ is fixed and equal to $n$, whereas the degeneracy of the first excited level at the end varies from instance to instance. We therefore first wanted to identify instances that have the highest number of first excited states at the end $s=1$ (remember that we always look at 3-SAT problems having exactly one solution, $r=1$). We calculated average degeneracies (averaged over $10000$ instances of random 3-SAT with $r=1$ solution) of the final hamiltonian $H(1)$ for $n=10$ and different $m$. The results are in Fig.~\ref{fig:spekter_n10}.   
We can see that the peak of the degeneracy moves to smaller energies with decreasing $m/n$. More importantly, the degeneracy of the first excited state (the height of the second bar) increases with decreasing $m/n$ as well. Therefore, to have as high degeneracy of the first excited state as possible, we have to have small ratio $m/n$. Note that nothing particular happens at the point of ``phase-transition'' at $m/n \approx 4-5$. From now on we will choose $m/n=3$, as such 3-SAT instances have high degeneracy of the first excited state. If this degeneracy increases exponentially with $n$ we have a fair chance that the gap $\Delta$ will also decrease exponentially. We therefore checked the scaling of the degeneracy of the first excited state of $H(1)$ with $n$ at a constant ratio $m/n=3$. 
\begin{figure}[!h]
\centerline{\includegraphics[height=3.3in,angle=-90]{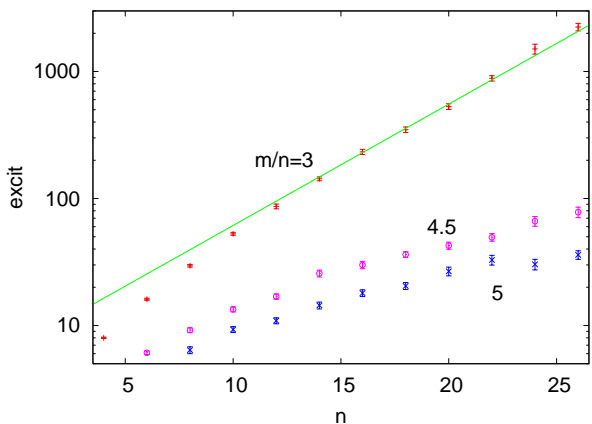}}
\caption{(Color online)~Degeneracy of the first excited state of the final $H(1)$ for different $n$ and $m/n$. Exponential growth can be seen for $m/n=3$, while for $m/n=5$ and $4.5$ (around the phase transition) the degeneracy is much smaller and the functional form of the growth can not be established.}
\label{fig:excit}
\end{figure}
Results are shown in Fig.~\ref{fig:excit}. We can see clear exponential growth for $m/n=3$ with the fitted exponential $\approx 8.2 e^{0.21 n}$. For larger $m/n$, i.e. around the phase transition, the growth is much slower and an exponential dependence can not be firmly established.
\par
Random 3-SAT instances with $m/n=3$ and exactly one solution are therefore our candidates for hard problems. For such instances the number of states violating only one clause (i.e. the degeneracy of the first excited state of $H(1)$) grows exponentially with $n$ and so the energy gap is expected to decrease exponentially and consequently running time to increase exponentially. What is the difficulty of such 3-SAT instances for classical algorithms? One could expect that classical algorithms will also have a hard time finding the solution as it is ``surrounded'' by exponentially many states that violate only one clause. To illustrate this we tested classical GSAT algorithm with random walk extension~\cite{Selman:92}. GSAT belongs to a group of incomplete algorithms for 3-SAT meaning that there is no a priori terminating condition. All such algorithms are variants of a local search. Concretely, in GSAT algorithm one variable is negated (flipped) at each step. The variable to negate is chosen so that the resulting state satisfies a maximum number of clauses. Note that GSAT algorithm can not prove unsatisfiability and is therefore suitable only for solvable instances. 
\begin{figure}[!h]
\centerline{\includegraphics[height=3.3in,angle=-90]{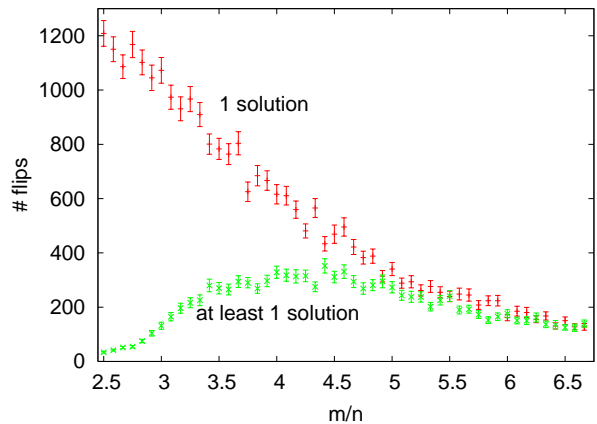}}
\caption{(Color online)~Running time (number of flips) for classical GSAT algorithm solving 3-SAT instances with exactly one solution (top points) and instances with at least one solution (bottom points). All is for $n=20$ and an average over $1000$ random 3-SAT instances is performed. Error bars show standard deviation at each point.}
\label{fig:gsat20}
\end{figure}
In Fig.~\ref{fig:gsat20} we show the average number of flips needed to find a solution for $n=20$ and different number of clauses $m$. Data shown is an average over $1000$ instances having exactly one solution, $r=1$, and for comparison also for problems with at least one solution, $r\ge 1$. We can see, that for problems with at least one solution we have a characteristic phase-transition dependence~\cite{Kirkpatrick:94,volumeAI} with the hardest instances at around $m/n \approx 4.2$. On the other hand, for $r=1$ the difficulty of problems grows with decreasing $m/n$. We see that such problems are actually {\em harder} than the problems at the phase transition. The same behaviour is expected also in other local search algorithms. On the other hand there are also complete methods for solving satisfiability problems. The most widely used is the so-called DPLL~\cite{dpll} algorithm and its derivations. It searches trough a solution tree and can both prove unsatisfiability or find a solution in a finite (exponential) number of steps. Preliminary results show~\cite{unpublished} that random 3-SAT instances with one solution are harder than instances at the phase-transition also for such complete algorithms.
\par
\begin{figure}[!h]
\centerline{\includegraphics[height=3.3in,angle=-90]{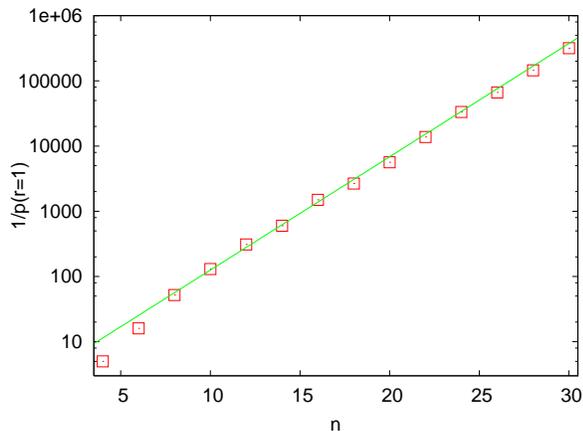}}
\caption{(Color online)~Frequency of 3-SAT formulas, i.e. the inverse probability to get such a formula, with exactly $r=1$ solution among random instances at $m/n=3$. Each point is an average over $100$ instances. Line is an exponential fit.}
\label{fig:prob_ground}
\end{figure}
Why have been such instances with one solution and small $m/n$ overlooked so far in a vast literature on phase-transition in random 3-SAT~\cite{Kirkpatrick:94,volumeAI,monasson:99,cocco:02}? The answer is very simple. They are exponentially rare among random 3-SAT instances (for small $m/n$ most have many solutions) and so they do not show up in the average behaviour that was usually studied. Nevertheless, as the computational complexity is defined in terms of a worst case performance, such problems are important. In Fig.~\ref{fig:prob_ground} we show how frequently one gets a 3-SAT problem with exactly one solution among randomly drawn 3-SAT problems at small $m/n=3$. The best fitting line in the figure is exponential $\approx 2.3 e^{0.4 n}$. The fact that 3-SAT with $r=1$ and small $m/n$ are exponentially rare makes their generation very time consuming as we have to solve very many instances before we arrive at the ``right one'' with exactly one solution.

\section{Scaling of running time}
\label{sec:gap}
Finally, after identifying 3-SAT instances that are expected to be hard for quantum adiabatic algorithm due to exponentially many states residing just above the ground state and after seeing that such instances are hard also for classical algorithms, we turn to the numerical calculation of the scaling of running time of quantum adiabatic algorithm for 3-SAT problems with $r=1$ and $m/n=3$. For each $n$ we generate $100$ such instances, find the position of the minimum of the energy gap and from the curvature at the gap determine the necessary running time according to the Landau-Zener formula (\ref{eq:LZ}). 
\begin{figure}[!h]
\centerline{\includegraphics[height=3.3in,angle=-90]{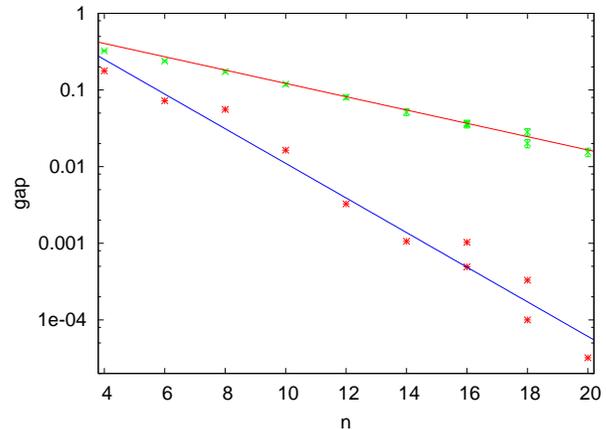}}
\caption{(Color online)~Dependence of the energy gap $\Delta$ on $n$ for 3-SAT with $m/n=3$ and $r=1$ solution. Top points are for the average gap and bottom for minimal gap (average over 100 instances). Two lines are exponential fits. For $n=16$ and $n=18$ we show the results of two independent runs to give the impression about fluctuations.}
\label{fig:gapnm3}
\end{figure}
In Fig.~\ref{fig:gapnm3} we show the dependence of the minimal gap $\Delta$ on the size $n$. Exponential fit gives dependence $\approx 0.9 e^{-0.2n}$ for the average $\Delta$ and $\approx 2 e^{-0.52 n}$ for a minimal $\Delta$ (out of $100$ instances). Clear exponential decrease can be seen over several orders of magnitude. Similar exponential behaviour can be seen also in the dependence of the running time $\tau_{\rm LZ}$ in Fig.~\ref{fig:timenm3}.
\begin{figure}[!h]
\centerline{\includegraphics[height=3.3in,angle=-90]{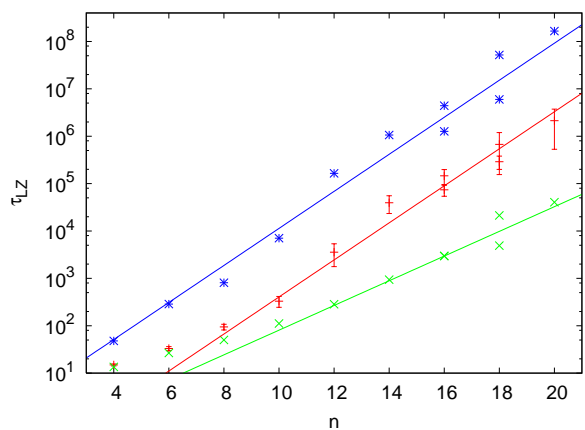}}
\caption{(Color online)~Dependence of running time (according to Landau-Zener formula) on $n$ for 3-SAT with $m/n=3$ and $r=1$ solution (same data as for Fig.~\ref{fig:gapnm3}). Top points are for a maximal gap, middle for the average and bottom for the median $\tau_{\rm LZ}$ (average over 100 instances). Exponential growth over $7$ orders of magnitude can be seen.}
\label{fig:timenm3}
\end{figure}
Exponential fits give scaling $\approx 1.4 e^{0.9n}$ for a maximal $\tau_{\rm LZ}$, $\approx 0.05 e^{0.9 n}$ for the average $\tau_{\rm LZ}$ and $\approx 0.2 e^{0.6 n}$ for a median time. We have therefore numerically established {\em exponential} growth of running time of the adiabatic quantum algorithm for 3-SAT with the problem size.  
\par
We should mention that if one looks at the scaling of e.g. energy gap for problems with one solution at larger $m/n$, say around the phase-transition, the functional dependence can not be firmly established. It looks like a power-law $\sim n^{-0.8}$, but the exponential dependence with a very small exponent (e.g. $\sim e^{-0.05 n}$) cannot be excluded. This is in agreement with a similar behaviour of degeneracies in Fig.~\ref{fig:excit}. It might be that the asymptotic regime of exponential dependence is not yet reached for small $n\le20$ amenable to numerical study. This is probably the reason why in previous studies no clear exponential time dependence could be identified.

\section{Conclusion}
We have numerically studied quantum adiabatic algorithm for 3-SAT problem. First, we identified a class of difficult random 3-SAT instances not known previously. These are instances with exactly one solution and small number of clauses (e.g. $m/n=3$). Such 3-SAT instances are exponentially rare, which is the reason they have not been observed so far. Nevertheless, as the computational complexity is concerned with the worst case performance, they determine the complexity of the algorithm. They have exponentially many assignments of variables that violate only one clause and thereby exponentially many states just above the ground state of the final Hamiltonian. Therefore, the energy gap for such instances decreases exponentially with the size of the problem and as a consequence, running time grows exponentially. This provides a firm numerical evidence that the usual quantum adiabatic algorithm for 3-SAT has exponential complexity. 
\par
In addition, such class of 3-SAT instances is expected to be difficult also for classical algorithms. In local search algorithms exponentially many ``fake'' solutions, violating only one clause, will effectively ``shadow out'' the real solution. In complete methods, like DPLL, a wrong assignment of a variable early in the search tree will cause large backtrackings. Becouse problems with small $m/n$ are under constrained, such wrong assignment will be very frequent which will make the search tree very large. As the performance of 3-SAT solving algorithms is important in many areas, further study of the behaviour of classical algorithms on this new class of 3-SAT problems in necessary.    
\par
I would like to thank Toma\v z Prosen and Gregor Veble for discussions.
Financial support by the grant P1-044 of the Ministry of Education, 
Science and Sports of Slovenia, and in part by the ARO grant (USA) DAAD 19-02-1-0086, as well as the Alexander von Humboldt Foundation are gratefully acknowledged.

\end{document}